\documentclass[a4paper,12pt]{article}
\usepackage{epsfig}
\usepackage{graphics}
\usepackage{graphicx}
\usepackage{indentfirst}
\usepackage[centerlast]{caption2}
\def\newpic#1{}

\begin{document}

\begin{center}
{\large \bf Asymptotic estimation of some multiple integrals and
the electromagnetic deuteron form factors at high momentum
transfer}

\vspace{1cm}

A.F. Krutov \footnote{E-mail: krutov@ssu samara.ru}

\vspace{5mm}
 \textsl{Department of General and
Theoretical Physics,}

\textsl{ Samara State University,}

\textsl{ Samara,}

\textsl{ 443011 Russia}

\vspace{1cm}

V.E. Troitsky \footnote{E-mail: troitsky@theory.sinp.msu.ru}

\vspace{5mm}

\textsl{D.V.~Skobeltsyn Institute of Nuclear Physics,}

\textsl{Moscow State University,}

\textsl{Moscow,}

\textsl{119992 Russia}

\vspace{1cm}

N.A. Tsirova \footnote{E-mail: ntsirova@ssu samara.ru}

\vspace{5mm}
 \textsl{Department of General and
Theoretical Physics,}

\textsl{ Samara State University,}

\textsl{ Samara,}

\textsl{ 443011 Russia}

\vspace{1cm}

(Dated: 10 September 2007)

\end{center}

\newpage

\begin{abstract}
A theorem about asymptotic estimation of multiple integral of a
special type is proved
 for the case when the integrand peaks at the
integration domain bound, but not at a point of extremum. Using
this theorem the asymptotic expansion of the electromagnetic
deuteron form factors at high momentum transfers is obtained in
the framework of two-nucleon model in both nonrelativistic and
relativistic impulse approximations. It is found that relativistic
effects slow down the decrease of deuteron form factors and
result in agreement between the relativistic asymptotics and
experimental data at high momentum transfers.
\end{abstract}

\vspace{1cm}


\newpage

\section{Introduction}

Recent advances in experimental investigations of hadron structure
arouse the interest in the theoretical study of the hadron
electromagnetic form factors at high momentum transfers (see e.g.
review \cite{GiG02} and references therein). In this connection
the JLab program of investigations on elastic electron-deuteron
scattering experiments at $Q^2\;\simeq\;$ 10 (GeV/c)$^2$
($Q^2=-q^2\;,\;q$ is transferred momentum) \cite{ArH05} attracts
considerable attention. There exists a hope that these JLab
experiments will help to determine the limits of application for
the two-nucleon model and to clarify the interplay between
nucleon-nucleon and quark approaches to the deuteron.

Using the asymptotic expansion presented in this paper we show
\cite{KrT07-1} that the momentum transfer region in the JLab
experiments is asymptotical for the deuteron considered as a
nucleon--nucleon system. That is why the study of the
electromagnetic deuteron form factors is interesting at
$Q^2\;\to\;\infty$.

The present work is devoted to the theoretical investigation of
deuteron form factors asymptotic behavior at high momentum
transfer in the framework of the two-nucleon model. The form
factors asymptotics is studied in accordance with the next points.

1. As a rule, calculation of the form factors asymptotic behavior
in the relativistic approaches reduces to the asymptotic
estimation of $n$-tuple integrals. In the relativistic approach
used in our work the deuteron form factors are expressed in terms
of double integrals where integrands  peak at the integration
domain bound, and corresponding point is not a point of extremum.
In this connection the theorem defining asymptotic expansion of
$n$-tuple integrals with such integrand is proven in our paper.

2. In general, high momentum transfers require relativistic
consideration. However we begin the consideration of the
asymptotic estimation of electromagnetic deuteron form factors
with the nonrelativistic case and non\-re\-la\-ti\-vis\-tic
impulse approximation at $Q^2\;\to\;\infty$ . This is due to the
facts that, at first, the nonrelativistic calculation is a less
complicated and, second, this calculation is important for
establishment of the role of the relativistic
effects.

3. The asymptotic expansion of the relativistic deuteron form
factors is calculated in the relativistic invariant impulse
approximation in a variant of instant form of Poincare-invariant
quantum mechanics (PIQM) developed in our papers previously
\cite{BaK95, KrT02, KrT03, KrT05, KrT07}. The relativistic
calculations are performed by analogy with nonrelativistic case.
It is shown that relativistic effects essentially slow down the
asymptotical decrease of the form factors.

4. It is found that obtained in the framework of the two-nucleon
model relativistic asymptotics coincides with the experimental
data.

This paper is organized as follows. Section 2 is devoted to the
proving of the central for this work theorem defining asymptotics
of multiple integrals of some special type. In Sec. 3 a brief
review of the formulas for the deuteron form factors in the
nonrelativistic and relativistic invariant impulse approximation
is given.  The deuteron form factors asymptotics is calculated in
nonrelativistic and relativistic impulse approximation with the
help of the proven theorem in Sec. 4. In Sec. 5 asymptotics of
the form factors is obtained for the deuteron wave functions in
the conventional representation as a discrete
su\-per\-po\-si\-ti\-on of Yukawa-type terms \cite{Mac01}. Sec. 6
contains the conclusions of this paper.

\section{Theorem on the asymptotic expansion of some multiple integrals in the
case when the maximal value of the integrand belongs to region
boundary}

In the following we will consider integrals of the kind:
\begin{equation}
F(\lambda)=\int\limits_{\Omega}f(\lambda,x)\exp[S(\lambda,x)]dx\;,
\label{intlap}
\end{equation}
where $\Omega$ is a domain in {\bf R}$^n$, $x=(x_1,...,x_n)$,
$\lambda$ is a large positive parameter. We will use following
definitions: $\partial\Omega$ is a bound of the domain $\Omega$,
$[\Omega]=\Omega\cup\partial\Omega$, the bound $\partial\Omega\in
C^{\infty}$ if in the vicinity of any point
$x^0\in\partial\Omega$ it can be specified by equation
$x_j=\varphi(x'),\ x'\in U',\
x'=(x_1,...,x_{j-1},x_{j+1},...,x_n)$,\ $U'$ is a neighborhood of
a point $x^{'\,0} $, and the function $\varphi (x') \in
C^{\infty}$ in $U'$.

Let us prove now a lemma and a theorem on the asymptotic
estimation of integrals in Eq.(\ref{intlap}).

L e m m a. {\it Let $S(\lambda,x)$ be a smooth function in the
$[\Omega]$, $f(\lambda,x)$ be a continuous function in the
$[\Omega]$, and $M(\lambda) \in C^1$,
$$
M(\lambda)=\sup\limits_{x\in[\Omega]}S(\lambda,x)<\infty\;,
$$
at some $\lambda_0>0$ the integral} (\ref{intlap}) {\it be
absolutely convergent:
$$
\int\limits_{\Omega}|f(\lambda_0,x)|\exp[S(\lambda_0,x)]dx<\infty\;,
$$
and the following conditions be fulfilled at
$\lambda\geq\lambda_0$:
\begin{equation}
\frac{\partial S(\lambda,x)}{\partial\lambda}\leq\frac{d
M(\lambda)}{d\lambda}\;, \label{uslovie1}
\end{equation}

\begin{equation}
|f(\lambda,x)|\leq C_1|f(\lambda_0,x)|\;. \label{uslovie2}
\end{equation}
Then at $\lambda\geq\lambda_0$ the following estimation is valid:}
\begin{equation}
|F(\lambda)|\leq C_2 e^{M(\lambda)}\;. \label{lemma}
\end{equation}

P r o o f. At $\lambda\geq\lambda_0$ the following estimations
are true:
$$
|F(\lambda)|\leq e^{M(\lambda)}\int\limits_{\Omega}
 e^{S(\lambda_0,x)-M(\lambda_0)} e^{S(\lambda,x)
-S(\lambda_0,x)-M(\lambda)+M(\lambda_0)}|f(\lambda,x)|dx\leq
$$
$$
\leq e^{M(\lambda)-M(\lambda_0)} \int\limits_{\Omega}
 e^{S(\lambda_0,x)+S(\lambda,x)
-S(\lambda_0,x)-M(\lambda)+M(\lambda_0)}|f(\lambda,x)|dx.
$$
From conditions (\ref{uslovie1}),(\ref{uslovie2}) we obtain the
inequality:
$$
|F(\lambda)|\leq C_1
 e^{M(\lambda)-M(\lambda_0)}\int\limits_{\Omega} e^{S(\lambda_0,x)}
|f(\lambda_0,x)|dx\leq C_2 e^{M(\lambda)}\;.
$$
Thus the statement (\ref{lemma}) of the lemma is proved.

Later we will consider function $S(\lambda,x)$ described in lemma
which has the maximal value in the point $x^0\in\partial\Omega$,
and $S(\lambda,x),\
\partial\Omega\in C^\infty$ in the vicinity
of $x^0$. This point is not the point of extremum, that means the
validity of the following conditions:
\begin{equation}
\frac{\partial S(\lambda,x^0)}{\partial n}\neq 0\;, \label{cond1}
\end{equation}
and matrix of coefficients $B$:
\begin{equation}
\left\|\frac{\partial^2S(\lambda,x^0)}{\partial\xi_i\partial\xi_j}
\right\|_{i,j=1}^{n-1}=B\;, \label{cond2}
\end{equation}
gives the negative determined quadratic form. In Eqs.
(\ref{cond1}), (\ref{cond2}) $\partial/\partial n$ specifies the
internal normal derivative $\vec{n}$ to the $\partial \Omega$, and
$\xi_1,...,\xi_{n-1}$ is an orthonormal basis in the tangential
to the $\partial\Omega$ plane $T\,\partial\Omega_{x^0}$ at the
$x^0$ point.

For convenience let us choose in the vicinity of point $x^0$ a
frame $y=(y_1,...,y_n)$, so that $x^0$ is the origin of
coordinate and the internal normal to $\partial\Omega$ coincides
with the last basis vector of the new coordinate system.
Functions $f,\ S$ in these coordinates we denote as $f^*,\ S^*$,
and $U^*$ is an image of $U$ (that is an image of a half-vicinity
of the point $x^0$). The equation for $\partial U^*$ in the
vicinity of the point $y=0$ can be written in the following way:
$$
y_n=\varphi(y'), \hspace{10mm} y'\in U',\hspace{10mm}
y'=(y_1,...,y_{n-1}),
$$
with $U'$ is a vicinity of the point $y'=0$, $\varphi(y')\in
C^\infty(U'),$ and at $y'\to 0,\varphi(y')=O(\mid y'\mid ^2)$.

T h e o r e m. {\it Let the following conditions be fulfilled:}

1$^\circ.\ f, S \in C([\Omega]).$

2$^\circ.$ $S(\lambda,x^0)$ {\it is maximal value of function}
$S(\lambda,x)\;,\;$ $x^0 \in \partial\Omega$, {\it and} $x^0$
{\it is not point of extremum.}

3$^\circ.\ f,\ S,\ \partial\Omega \in C^{\infty}$ {\it in the
vicinity of the point} $x^0$.

4$^\circ.$ {\it The Taylor expansion of functions $S^*$ and $f^*$
in the vicinity of point $x^0$ satisfy the following relations:}
\begin{equation}
f^*(\lambda, y)=f^*(\lambda, 0)[1+o(1)]\;, \label{f*ly}
\end{equation}
\begin{equation}
 S^*(\lambda, y',\varphi(y'))-S^*(\lambda, 0)=\frac{1}{2}\langle Ay',\
y'\rangle+O(\mid y'\mid ^3)\;, \label{S*ly}
\end{equation}
{\it the matrix
$A=\left\|\frac{\partial^2S^*(\lambda,0)}{\partial y_i\partial
y_j} \right\|_{i,j=1}^{n-1},$ angle brackets denote bilinear
form: $\langle x,\ y\rangle=x_1y_1+x_2y_2+...+x_ny_n$}.

{\it Then at} $\lambda\to\infty$ {\it the following asymptotic
expansion is valid:}
\begin{equation}
F(\lambda)\sim\exp[S(\lambda,x^0)]\sum\limits_{k=0}^{\infty}a_k(\lambda)\;.
\label{ocenka}
\end{equation}
{\it The way to calculate coefficients $a_k(\lambda)$ will be
determined later.}

P r o o f. Let us divide the integral (\ref{intlap}) into two
integrals. Integration domain of the first one is the
half-vicinity $U$ of the point $x^0$, and integration domain of
the second one is a remainder of integration domain of the
original integral. It is easily shown by the proven lemma that
the second integral is exponentially small as compared with
$\exp[S(\lambda,x^0)]$. So we will estimate asymptotically the
first integral only.

In the expansion of the function $S^*(\lambda,y',\varphi(y'))$ in
line with the condition (\ref{S*ly}) linear components are absent,
because the point $y'=0$ is a point of maximum of the function
$S^*(\lambda,y',\varphi(y'))$ in the region $U'$.

Let us choose $U$ in accordance with inequalities
$\varphi(y')\leq y_n\leq\delta,\ \delta>0$ at $y \in U^*$. Then
we can represent the integral (\ref{intlap}) within exponentially
decreasing terms:

\begin{equation}
F(\lambda)=\int\limits_{U^*}f^*(\lambda,
y)\exp[S^*(\lambda,y)]dy\;. \label{idok}
\end{equation}
Let us rewrite integral in Eq. (\ref{idok}) in the following way:
$$
F(\lambda)=\int\limits_{U'}\Phi(\lambda, y')dy'\;,
$$
with
\begin{equation}
\Phi(\lambda, y')=\int\limits_{\varphi(y')}^{\delta}
\exp[S^*(\lambda,y)]f^*(y)dy_n\;. \label{intfi}
\end{equation}
The integral (\ref{intfi}) is one-dimensional, and the function
$S^*(\lambda, y)$ reaches extremum on the boundary
$y_n=\varphi(y')$. Asymptotic expansion of this integral can be
found through integration by parts. After $N+1$ integration we
obtain the sequence:
$$
\Phi(\lambda,y')=\sum\limits_{k=0}^{N}M^k\left[\frac{f^*(\lambda,y)}{S^{*'}(\lambda,y)}\right]
\left.\exp[S^*(\lambda,y)]\right|_{\varphi(y')}^{\delta}-\int\limits_{\varphi(y')}^{\delta}
{M^N\left[\frac{f^*(\lambda,y)}{S^{*'}(\lambda,y)}\right]'}\exp[S^*(\lambda,y)]dy_n\;,
$$
with $M^0$ is a unit operator and
$$
M^k=-\frac{1}{S^{*'}(\lambda,y)}\frac{d^k}{dy^k_n}\;.
$$
The substitution of $y_n=\varphi(y')$ provides the main
contribution to the asymptotics, the value of $y_n=\delta$ is
exponentially small as compared with the previous. Further
integration  under these conditions gives the following expansion
for the function (\ref{intfi}):
$$
\Phi(\lambda,y')=-\exp[S^*(\lambda,y',\varphi(y'))]\sum\limits_{k=0}^\infty
M^k\left[\frac{f^*(\lambda,y',\varphi(y'))}{S^{*'}(\lambda,y',\varphi(y'))}\right]\;.
$$
So
\begin{equation}
F(\lambda)=-\sum\limits_{k=0}^\infty
\int\limits_{U'}\exp[S^*(\lambda,y',\varphi(y'))]
M^k\left[\frac{f^*(\lambda,y',\varphi(y'))}{S^{*'}(\lambda,y',\varphi(y'))}\right]dy'
\label{F}
\end{equation}
The point $y'=0$ is an internal point of maximum of the integrand
in the expression (\ref{F}). Functions
$S^*(\lambda,y',\varphi(y'))$ and $f^*(\lambda,y',\varphi(y'))$
satisfy the conditions of lemma (\ref{uslovie1}) and theorem
(\ref{f*ly}), (\ref{S*ly}), therefore we can apply a formula for
asymptotic estimation of the n-tuple Laplas integrals
\cite{Fed77} and obtain corresponding asymptotic expansion
(\ref{ocenka}) of the integral in Eq. (\ref{F}). Thus the theorem
is proven.

Note, that in the general case it is rather difficult to write a
compact formula for the coefficients $a_k(\lambda)$ in Eq.
(\ref{ocenka}). They can be obtained such kind of way in any
specific cases. As an example, these coefficients will be
obtained and written out explicitly for double integrals in the
consideration of the asymptotic estimation of electromagnetic
deuteron form factors in Sec. 4. Here we write out only the first
asymptotic term  from Eq. (\ref{ocenka}) in the $x$ variables:

\begin{equation}
F(\lambda)\; \sim\;
-(2\pi)^{\frac{n-1}{2}}\exp[S(\lambda,x^0)]\left(\frac{\partial
S(\lambda,x^0)}{\partial n}\right)^{-1}|\det
B|^{-\frac{1}{2}}f(\lambda,x^0)\;, \label{gch}
\end{equation}
where $\vec n\;,\;B$ are defined by conditions (\ref{cond1}) and
(\ref{cond2}).

\section{Electromagnetic deuteron form factors in the \\
nonrelativistic and relativistic impulse approximation}

In the nonrelativistic impulse approximation known formulas for
electromagnetic deuteron form factors can be rewritten in the
following way \cite{JaM72}:
$$
G^{NR}_C(Q^2) = \sum_{l,l'}\int\,k^2\,dk\,k'\,^2\,dk'\, u_l(k)\,
\tilde g^{ll'}_{0C}(k\,,Q^2\,,k')\, u_{l'}(k')\;,
$$
$$G^{NR}_Q(Q^2) =
\frac{2\,M_d^2}{Q^2}\,\sum_{l,l'}\int\,k^2\,dk\,k'\,^2\,dk'\,
u_l(k)\,\tilde g^{ll'}_{0Q}(k\,,Q^2\,,k')\,u_{l'}(k')\;,
$$
\begin{equation}\label{GqGNIP}
G^{NR}_M(Q^2) =-\,M_d\,\sum_{l,l'}\int\,k^2\,dk\,k'\,^2\,dk'\,
u_l(k)\,\tilde g^{ll'}_{0M}(k\,,Q^2\,,k')\, u_{l'}(k')\;.
\end{equation}
Here $u_l(k)$ are the deuteron wave functions in momentum
representation, $l\;,\;l'=$ 0,2 are orbital angular momenta,
$\tilde g^{ll'}_{0i}(k\,,Q^2\,,k')\;,\;i=C,Q,M$ are
nonrelativistic free two-particles charge, quadrupole and
magnetic dipole form factors, $M_d$ is the deuteron mass.
Formulas for $\tilde g^{ll'}_{0i}$ are given in \cite{KrT07}.

Let us discuss briefly possible types of the model deuteron wave
functions. There are several classes of the deuteron wave
functions: obtained with microscopic model Hamiltonians of the
$NN$-interaction in the non-relativistic nuclear physics (for
example, see \cite{Mac01}), deduced from scattering amplitudes in
the Bethe-Salpeter approach and its various quasipotential
reductions (see \cite{StG97}), wave functions of the
Poincare-invariant quantum mechanics (as an example see wave
functions in the instant form of PIQM \cite{BaK95, KrT02, KrT03,
KrT05, KrT07}), and also wave functions calculated in the various
statements of inverse scattering problems \cite{Tro94,
KrM97,ShM03}. But independently of the method any wave function
can be represented as the following Laguerre polynomial expansion
\cite{ShM03}:
\begin{equation}
u_l(k)=\sum\limits_{m=0}^{\infty} a_{lm}\sqrt{\frac{2m!}{\Gamma
(m+l+3/2)}}\; r_0^{l+\frac{3}{2}}\;k^l\;
L_{m}^{l+\frac{1}{2}}(r_0^2 k^2)\; e^{-\frac{r_0^2 k^2}{2}}\;
\label{funcu}
\end{equation}
or in the coordinate representation:
\begin{equation}
u_{l}(r) = \sum\limits_{m=0}^{\infty} (-1)^m
a_{lm}\sqrt{\frac{2m!}{r_0\;\Gamma (m+l+3/2)}}\;
\left(\frac{r}{r_0}\right)^{l+1}\;
L_{m}^{l+\frac{1}{2}}\left(\frac{r^2}{r_0^2}\right)\;
e^{-\frac{r^2}{2r_0^2}}\;, \label{funcukoord}
\end{equation}
here $L^{l+1/2}_{m}(x)$ are generalized Laguerre polynomials,
$\Gamma (x)$ is an Euler gamma function, the dimensional parameter
$r_0$ can be related to the deuteron matter radius (see Sec. 5).

The wave function representation as a Laguerre polynomial
expansion (\ref{funcu}) is very useful for the calculation of the
asymptotic behavior of the form factors. However, one can avoid
such representation and obtain the asymptotic expansion directly
for the initial wave function.

Generally, at high transferred momentum it is necessary to take
into account relativistic corrections in the electromagnetic
deuteron structure. In our paper relativistic description of the
deuteron is constructed in the framework of instant form of
Poincare-invariant quantum mechanics (PIQM), developed by authors
previously \cite{BaK95,KrT02,KrT03,KrT05,KrT07}. In this approach
we present electromagnetic deuteron form factors by analogy with
nonrelativistic case (\ref{GqGNIP}). Corresponding formulas in
the relativistic impulse approximation were obtained in our paper
\cite{KrT03}:
$$
G^R_C(Q^2) = \sum_{l,l'}\int\,d\sqrt{s}\,d\sqrt{s'}\,
\varphi_l(s)\, g^{ll'}_{0C}(s\,,Q^2\,,s')\, \varphi_{l'}(s')\;,
$$
$$
G^R_Q(Q^2) =
\frac{2\,M_d^2}{Q^2}\,\sum_{l,l'}\int\,d\sqrt{s}\,d\sqrt{s'}\,
\varphi_l(s)\,g^{ll'}_{0Q}(s\,,Q^2\,,s')\,\varphi_{l'}(s')\;,
$$
\begin{equation}
\label{GqGRIP}
G^R_M(Q^2)=-\,M_d\,\sum_{l,l'}\int\,d\sqrt{s}\,d\sqrt{s'}\,
\varphi_l(s)\,g^{ll'}_{0M}(s\,,Q^2\,,s')\, \varphi_{l'}(s')\;,
\end{equation}
where $\varphi_l(s)$ are the deuteron wave functions in sense of
PIQM, $g^{ll'}_{0i}((s\,,Q^2\,,s'),$ $i=C,Q,M$ are relativistic
free two-particles charge, quadrupole and magnetic dipole form
factors. Formulas for free form factors are given in \cite{KrT07}.

The deuteron wave functions in sense of PIQM are solutions of
eigenvalue problem for a mass squared operator for the deuteron
(see, e.g. \cite{BaK95}): $ \hat M^2_d\, |\psi\rangle =
M^2_d\,|\psi\rangle. $ An eigenvalue problem for this operator is
coincident with the nonrelativistic Schr\"odinger equation within
a second order on deuteron binding energy
${\varepsilon_d^2}/({4M})$, the value of which is small ($M$ is
an averaged nucleon mass). So the deuteron wave functions in
sense of PIQM differ from nonrelativistic wave functions by
conditions of normalization only. In the relativistic case the
wave functions are normalized with relativistic density of states:
\begin{equation}
\sum_{l=0,2}\,\int_0^\infty\varphi^2_l(k)\,\frac{dk}{2\sqrt{k^2 +
M^2}}=1\;,\quad
 \varphi_l(k) = \sqrt[4\ ]{s}\,k\,u_l(k)\;,\quad
s=4(k^{2}+M^{2})\;.\label{relnorm}
\end{equation}

Nonrelativistic formulas Eq. (\ref{GqGNIP}) can be obtained from
relativistic ones (\ref{GqGRIP}) in the nonrelativistic limit.
This limiting procedure can be performed in the most natural way
in the instant form of PIQM. The reason is that in papers
\cite{BaK95,KrT02,KrT03,KrT05,KrT07} we have constructed the
successful formalism of the instant form of PIQM. In the case of
other forms of PIQM (point and front forms) the obtaining of
nonrelativistic limit is much more difficult.

For obtaining the asymptotic form factors behavior at high
transferred momentum in the nonrelativistic and relativistic
cases it is necessary to estimate asymptotically double integrals
(\ref{GqGNIP}) and (\ref{GqGRIP}) at $Q^2\to\infty$. Notice that
integrands reach its maximum value at the integration domain
bound, and this point is not a point of extremum. In the previous
Section the theorem defining asymptotics of $n$-tuple integrals
of such kind was proven.

\section{Asymptotic expansion of the deuteron form factors}

We start asymptotic expansion of the deuteron form factors from
the nonrelativistic case. It is caused by the simplicity of the
nonrelativistic formulas, so the calculation of the asymptotics
is more clear. In what follows the relativistic calculation will
be presented analogous to the nonrelativistic one, although more
combersome. Moreover nonrelativistic calculation is interesting
because nonrelativistic formulas for the form factors
(\ref{GqGNIP}) are conventional, that is why its correct
asymptotic expansion has universal significance. Let us emphasize
also that the relativistic expressions for form factors and,
therefore, their asymptotic estimations depend on the choice of
the method of relativisation of the two-nucleon model.
Nonrelativistic calculation is also of interest because it helps
to clarify the role of relativistic effects in the
electromagnetic structure of the deuteron at the asymptotical
momentum transfers.

As we have seen in Sec. 3, the deuteron form factors in the
nonrelativistic impulse approximation can be represented by
double integrals (\ref{GqGNIP}). We will find its asymptotic
expansion using the theorem of Section 2 and use as an example
 the asymptotics of the charge form factor. We shall estimate
only the $l = l' = 0$ term in the sum (\ref{GqGNIP}) because the
asymptotics of the other terms of form factors (\ref{GqGNIP}) can
be derived analogously.

Let us rewrite the corresponding $l = l' = 0$ term of the charge
form factor (\ref{GqGNIP}) using Eq. (\ref{funcu}):
$$
\int\,\tilde g^{00}_{0C}(k\,,Q^2\,,k')\,\exp\left[S(k,k')\right]
\,k^2\,dk\,k'\,^2\,dk'\;\times
$$
\begin{equation}
\times\left(\sum\limits_{m} a_{0m}\sqrt{\frac{2m!}{\Gamma
(m+3/2)}}\; r_0^{\frac{3}{2}}\; L_{m}^{\frac{1}{2}}(r_0^2
k^2)\right)\left(\sum\limits_{m} a_{0m}\sqrt{\frac{2m!}{\Gamma
(m+3/2)}}\; r_0^{\frac{3}{2}}\; L_{m}^{\frac{1}{2}}(r_0^2
k'\,^2)\right)\;. \label{Gcnew}
\end{equation}
We have denoted in (\ref{Gcnew}):
\begin{equation}
S(k,k') = -\frac{r_0^2}{2} \;(k^2+k'^2)\;.
 \label{S}
\end{equation}
The expression for $\tilde g^{00}_{0C}(k\,,Q^2\,,k')$ is commonly
accepted (see, e.g., \cite{KrT07}):
\begin{equation}
\tilde g^{00}_{0C}(k, Q^2, k') = \frac{1}{k\,k'\,Q}
\left[\theta\left(k' - \left|k - \frac{Q}{2}\right|\right) -
\theta\left(k' - k - \frac{Q}{2}\right)\right]
\left(G^p_E(Q^2)+G^{n}_E(Q^2)\right),\label{tg00}
\end{equation}
$G^{p,n}_E(Q^2)$ are electric form factors of proton and neuteron
respectively, $\theta(x)$ is a step function.

In the case under consideration the space dimension $n=2,\
(x_1,x_2)=(k,k'),\ \lambda = Q^2$ is a large positive parameter.
Integration domain is determined by $\theta$-functions in Eq.
(\ref{tg00}) and shown in Fig.1. The location of the point of
maximal value of the function $S$ can be obtained by analysis of
(\ref{S}) and (\ref{tg00}):
$(k^0,k'\,^0)=(\frac{Q}{4},\frac{Q}{4})$.

\begin{figure}
\begin{center}
\includegraphics[scale=1]{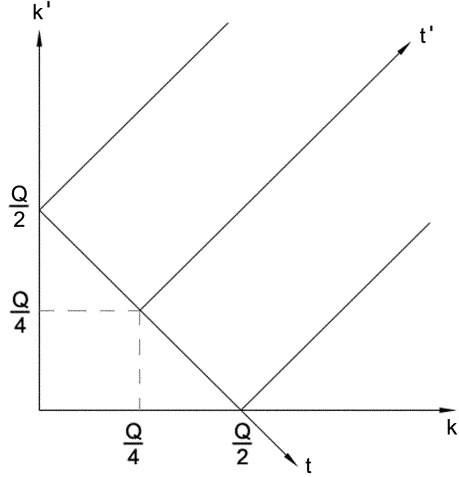}
\caption{The integration domain, location of the point of maximal
value, and transition to the new variables for the
nonrelativistic case} \label{dff}
\end{center}
\end{figure}

Let us perform the transition to the new basis as we have
descripted before. We perform the shift of the origin of
coordinates to the point of maximal value of the function $S$.
Then we rotate the obtained frame for the internal normal to the
boundary in the new origin to be coincident with the last basis
vector of the new frame. This procedure is illustrated in Fig.1.

In the other words we perform the transition to the new variables
in Eq. (\ref{Gcnew}):
\begin{equation}
k=\frac{1}{\sqrt{2}}\left(t' + t\right) + \frac{Q}{4}\;, \quad
k'=\frac{1}{\sqrt{2}}\left(t' - t\right) + \frac{Q}{4}\;.
\label{tt'}
\end{equation}
At this transformation function $S(k,k')$ gets dependence on large
parameter $Q^2$:
\begin{equation}
S^*(Q^2,t,t') =
-\frac{r_0^2}{2}\left(t^2+t'\;^2+\frac{Q}{\sqrt{2}}\;t'+\frac{Q^2}{8}\right)
\label{s}
\end{equation}
Functions $S(Q^2,t,t')\;,g^{00}_{0C}(k\,,Q^2\,,k')$, and the
boundary of the integration domain satisfy the conditions of
theorem 1$^\circ$, 3$^\circ$, 4$^\circ$. The location of the
point, that satisfies the conditions (\ref{cond1}),
(\ref{cond2}), can be obtained by a simple analysis of the
function (\ref{s}): $(t^0,t'\,^0)=(0,0)$. Let us show that this
point satisfies the conditions 2$^\circ$ of the theorem.

It is obvious, that in the point of maximal value
$$
\left.\frac{\partial S}{\partial n}=\frac{\partial S^*}{\partial
t'}\right|_{(t,t')=(0,0)}= -\frac{r_0^2\;Q}{2 \sqrt{2}}\;\ne\;
0\;.
$$
So the condition (\ref{cond1}) is satisfied.

Let us calculate now the $B$ matrix from the condition
(\ref{cond2}). In our case the tangent to the domain of
integration boundary vector in the point of maximal value is
$\vec{\xi}=(1/\sqrt{2},-1/\sqrt{2})$, i.e. the $B$ matrix is a
number:
\begin{equation}
\left.\frac{\partial^2 S}{\partial\xi^2}=\frac{\partial^2
S^*}{\partial t^2}\right|_{(t,t')=(0,0)}\;= -r_0^2\;<\;0\;.
 \label{B11}
\end{equation}
We note, that $B$ is negative-definite, i.e. the point
$(t^0,t^0\,') = (0,0)$ is really the point of maximal value. So
the point (0,0) satisfies the condition 2$^\circ$ of the theorem.

So integral (\ref{Gcnew}) satisfies the requirements of the
theorem proven in the Sec. 2. Therefore we can apply the
asymptotic formula (\ref{ocenka}).

Calculating by analogy the other terms of the sum (\ref{GqGNIP})
we obtain asymptotic expansions of deuteron form factors in the
nonrelativistic impulse approximation:
\begin{equation}
G_i^{NR}(Q^2)\sim e^{-\frac{r_0^2
Q^2}{16}}\sum\limits_{m=0}^\infty
\frac{h_{2m}^{NR}}{(2m)!}\;\Gamma(m+\frac{1}{2}), \label{ryadner}
\end{equation}
\begin{equation}
h_{2m}^{NR}=\left.\sum\limits_{k=0}^\infty\frac{1}{r_0^{2m+2k+3}}\sum\limits_{p=0}^k
b_{kp} \left(\frac{2\sqrt{2}}{Q}\right)^{p+k+1}
\frac{\partial^{2m}}{\partial
t^{2m}}f_i^{NR\;(k-p)}(t,Q^2,0)\right|_{t=0}, \label{koefner}
\end{equation}
$$
b_{k0}=1,\;b_{kp}=b_{k-1\;p}-(k+p-1)b_{k-1\;p-1},\;b_{kk}=(-1)^k(2k-1)!!\;,
$$
$$
f_i^{NR}(t,Q^2,t')=A_i\sum\limits_{l,l'=0,2}k^{l+2}\,k'\,^{l'+2}\,
\tilde{u}_l\left(k\right)\;\tilde{g}^{ll'}_{0i}(t,Q^2,t')\;\tilde{u}_{l'}
\left(k'\right)\;,
$$
with $k=k(t,Q^2,t')\,,k'=k'(t,Q^2,t')$, variables $t,\;t'$ are
denoted in (\ref{tt'}), $i=C,Q,M$,
$\;A_C=1,\;A_Q=2M_d^2/Q^2\,,\;A_M=-M_d$,
$$
f_i^{NR\;(m)}(t,Q^2,t')=\frac{\partial^m}{\partial
t'\,^{m}}f_i^{NR}(t,Q^2,t')\;.
$$
$\tilde{u}_{l,l'}$ is defined by equalities:
\begin{equation}
u_0(k)=\tilde{u}_0(k)\;e^{-\frac{r_0^2
k^2}{2}}\;,\;u_2(k)=\tilde{u}_2(k)\;k^2\;e^{-\frac{r_0^2
k^2}{2}}\;. \label{utildeu}
\end{equation}
Let us perform now the calculation of the relativistic
asymptotics of deuteron form factors. To estimate asymptotically
integrals (\ref{GqGRIP}) we proceed analogously to
nonrelativistic case, i.e. we use relativistic analogs of
corresponding nonrelativistic formulas (\ref{Gcnew})-(\ref{B11}).
Now the free relativistic charge form factor in (\ref{GqGRIP}) at
$l = l' = 0$ is given in Ref. \cite{KrT07}:
$$
g^{00}_{0C}(s, Q^2, s') = R(s, Q^2, s')\,Q^2 \left[\,(s + s' +
Q^2)\left(G^p_E(Q^2)+G^{n} _E(Q^2)\right)g^{00}_{CE} + \right.
$$
\begin{equation}
\left. + \frac{1}{M}\xi(s,Q^2,s')
\left(G^p_M(Q^2)+G^{n}_M(Q^2)\right)g^{00}_{CM}
\right]\;,\label{g00}
\end{equation}
$G^{p,n}_{E,M}(Q^2)$ are electric and magnetic form factors of
proton and neutron respectively,
$$
g^{00}_{CE} = \left(\frac{1}{2}\cos\omega_1\cos\omega_2 +
\frac{1}{6}\sin\omega_1\sin\omega_2\right)\;,\quad g^{00}_{CM} =
\left(\frac{1}{2}\cos\omega_1\sin\omega_2 -
\frac{1}{6}\sin\omega_1\cos\omega_2\right)\;,
$$
$$
R(s, Q^2, s') = \frac{(s+s'+Q^2)}{\sqrt{(s-4M^2) (s'-4M^2)}}\,
\frac{\vartheta(s,Q^2,s')}{{[\lambda(s,-Q^2,s')]}^{3/2}}
\frac{1}{\sqrt{1+Q^2/4M^2}}\;,
$$
$$
\xi(s,Q^2,s')=\sqrt{ss'Q^2-M^2\lambda(s,-Q^2,s')}\;,
$$
$\omega_1$ and $\omega_2$ are angles of the Wigner spin rotation,
$$
\omega_1 =
\arctan\frac{\xi(s,Q^2,s')}{M\left[(\sqrt{s}+\sqrt{s'})^2 +
Q^2\right] + \sqrt{ss'}(\sqrt{s} +\sqrt{s'})}\>,
$$
\begin{equation}
\omega_2 = \arctan\frac{ \alpha (s,s') \xi(s,Q^2,s')} {M(s + s' +
Q^2) \alpha (s,s') + \sqrt{ss'}(4M^2 + Q^2)}\>, \label{omega}
\end{equation}
where $\alpha (s,s') = 2M + \sqrt{s} + \sqrt{s'}$,
$\vartheta(s,Q^2,s')= \theta(s'-s_1)-\theta(s'-s_2)$, $\theta$ is
a step function, $\lambda(a,b,c)=a^2+b^2+c^2-2(ab+ac+bc)$,
$$
s_{1,2}=2M^2+\frac{1}{2M^2} (2M^2+Q^2)(s-2M^2) \mp \frac{1}{2M^2}
\sqrt{Q^2(Q^2+4M^2)s(s-4M^2)}\;.
$$

To obtain relativistic asymptotic expansion we also perform
transition to the new basis (shift and rotation). The function
$S$ and the boundary of the integration domain differ from
nonrelativistic ones, so it is necessary to perform a special
analysis. In other words, instead of change of variables
(\ref{tt'}) we perform  the following replacement:
\begin{equation}
s=\frac{1}{\sqrt{2}}\left(t' + \frac{t}{Q}\right) + 2 M^2 +
M\sqrt{Q^2 + 4 M^2}\;,\quad s'=\frac{1}{\sqrt{2}}\left(t' -
\frac{t}{Q}\right) + 2 M^2 + M\sqrt{Q^2 + 4 M^2}\;. \label{tt'rel}
\end{equation}

Then we obtain asymptotic expansion of the relativistic deuteron
form factors by analogy with nonrelativistic case:
\begin{equation}
G^R_i(Q^2)\sim
e^{-\frac{r_0^2}{4}(M\sqrt{Q^2+4M^2}-2M^2)}\sum\limits_{m=0}^\infty
\frac{h^R_{2m}}{(2m)!} \Gamma (m+\frac{1}{2}), \label{ryadrel}
\end{equation}
\begin{equation}
h^R_{2m}=\left.\sum\limits_{k=0}^\infty\sum\limits_{p=0}^{2m}\frac{(-1)^p}{Q^{p+m-\frac{1}{2}}}
\;(2p-1)!!\;
C^{2m}_{2p}\;M^{m-p+\frac{1}{2}}\;\frac{2^{\frac{5}{2}k+4m-2p+\frac{9}{2}}}{r_0^{2k+2m-2p+3}}\;
\frac{\partial^{2m-2p}}{\partial t^{2m-2p}}
f_i^{R\;(k)}(t,Q^2,\varphi(t))\right|_{t=0}, \label{koefrel}
\end{equation}

$$
f_i^R(t,Q^2,t')=A_i\sum\limits_{l,l'=0,2}\tilde{u}_l(k)
\;g^{ll'}_{0i}(t,Q^2,t')\;
\tilde{u}_{l'}(k')\frac{\left(s/4-M^2\right)^{\frac{l+1}{2}}
\left(s'/4-M^2\right)^{\frac{l'+1}{2}}}{\sqrt[4]{s\,s'}},
$$
$$
f_i^{R\;(m)}(t,Q^2,t')=\frac{\partial^m}{\partial
t'\,^{m}}f_i^{R}(t,Q^2,t')\;.
$$
Functions $k=k(s)\,,k'=k'(s')$ are specified in Eq.
(\ref{relnorm}), $s=s(t,Q^2,t')\,,s'=s'(t,Q^2,t')$, variables
$t,\;t'$ are denoted in (\ref{tt'rel}), $C^{2m}_{2p}$ are the
binomial coefficients.

Asymptotic expansions (\ref{ryadner}) and (\ref{ryadrel}) are
convergent power series in inverse degrees of the parameter $Q$
with known coefficients. The asymptotic expansion of this type is
obtained in this work for the first time.

One can see from formulas (\ref{ryadner}) and (\ref{ryadrel}),
that relativistic corrections change the behavior of form factors
at high momentum transfer. In particular, exponential multiplier
index is $Q^2$ in the nonrelativistic case, but in the
relativistic case it is $Q$ at $Q^2\to\infty$. It seems to be a
general feature of our relativistic approach to the description
of composite systems, in particular, we have obtained the similar
result in consideration of asymptotic behavior of the pion form
factor in the composite quark model \cite{KrT98}.

\section{Asymptotics of the form factors for the conventional wave functions representation}

In this Section we represent the obtained asymptotical expansions
(\ref{ryadner}) and (\ref{ryadrel}) in terms of initial wave
functions in the left side of Eqs. (\ref{funcu}),
(\ref{relnorm}). For this representation it is necessary to
replace functions $\tilde{u}_l(k)$ by functions ${u}_l(k)$ in
(\ref{ryadner}) and (\ref{ryadrel}) using (\ref{relnorm}),
(\ref{utildeu}). Keeping the main term on $1/Q$ in asymptotic
expansions (\ref{ryadner}) and (\ref{ryadrel}) one can obtain the
next asymptotic formulas in terms of functions $u_l(k)$ and
$\varphi_l(s)$ from (\ref{funcu}), (\ref{relnorm}):
\begin{equation}
G_i^{NR}(Q^2)=\left.-A_i\frac{4\sqrt{\pi}}{r_0^3Q}
\sum\limits_{l,l'=0,2}k^2\;k'\,^{2}
u_l(k)\;\tilde{g}^{ll'}_{0i}(t,Q^2,t')\;u_{l'}(k')\right|_{{{t=0}\atop
{t'=0}}}, \label{anywfner}
\end{equation}
\begin{equation}
G^R_i(Q^2)=\left.-A_i\frac{8\;\sqrt{\sqrt{2}\pi M}}{r_0^3\sqrt{Q}}
\sum\limits_{l,l'=0,2}
\frac{\varphi_l(s)\;g^{ll'}_{0i}(t,Q^2,t')\;\varphi_{l'}
(s')}{\sqrt[4]{s\,s'}}\right|_{{{t=0}\atop {t'=0}}}\;,\quad
i=C,Q,M\;. \label{anywfrel}
\end{equation}
Let us note, that similar asymptotic representation can be
obtained for any finite number of terms in asymptotic expansions
(\ref{ryadner}), (\ref{ryadrel}).

In the modern calculations the deuteron wave functions are usually
represented as a discrete superposition of Yukawa-type terms (see,
e.g., \cite{Mac01}):
\begin{equation}
u_0(k)=\sqrt{\frac{2}{\pi}}\sum\limits_j
\frac{C_j}{(k^2+m_j^2)},\;\quad
u_2(k)=\sqrt{\frac{2}{\pi}}\sum\limits_j
\frac{D_j}{(k^2+m_j^2)},\; \label{paru}
\end{equation}
or in the coordinate representation:
$$
u_0(r) = \sum\limits_j {C_j}{\exp\left(-m_j\,r\right)}\;,
$$
\begin{equation}
u_2(r) = \sum\limits_j {D_j}{\exp\left(-m_j\,r\right)} \left[1 +
\frac{3}{m_j\,r} + \frac{3}{(m_j\,r)^2}\right]\;,
\label{parukoord}
\end{equation}
$$
m_j = \alpha + m_0\,(j-1)\;,\quad \alpha =
\sqrt{M\,\left|\varepsilon_d\right|}\;.
$$
Coefficients $C_j,\,D_j$, maximal value of the index $j$ and
$m_0$ are determined by the best fit of corresponding solution of
Schr\"{o}dinger equation.

The deuteron wave function analytical form (\ref{parukoord})
results in the right behavior of the wave functions at large
distances:
\begin{equation}
u_{0}(r)\;\sim\;\exp(-\,\alpha\,r)\;,\quad
u_{2}(r)\;\sim\;\exp(-\,\alpha\,r)\left(1 + \frac{3}{(\alpha\,r)}
+ \frac{3}{(\alpha\,r)^2}\right)\;. \label{rinfty}
\end{equation}
The deuteron wave functions behavior at small distances:
\begin{equation}
u_0(r)\;\sim\;r\;,\quad u_2(r)\;\sim\;r^3\;, \label{atorigin}
\end{equation}
is provided by imposing the following conditions on coefficients
$C_j$ and $D_j$:
\begin{equation}
\sum\limits_j {C_j} = 0\;,\quad
\sum\limits_j {D_j} =
\sum\limits_j {D_j}{m_j^2} = \sum\limits_j \frac{D_j}{m_j^2} =
0\;. \label{cond}
\end{equation}

Let us substitute the wave functions (\ref{paru}) to
(\ref{anywfner}) and (\ref{anywfrel}), and then obtain the first
asymptotic terms of the nonrelativistic deuteron form factors:

\begin{equation}
G_C^{NR}\sim
\frac{1}{Q^8}\frac{2^{16}}{\sqrt{\pi}r_0^3}\left[\sum\limits_{j}C_{j}m_j^2\right]^2
\left(G_E^p(Q^2)+G_E^n(Q^2)\right)\;, \label{gnrc}
\end{equation}

\begin{equation}
G_Q^{NR}\sim
3\,M_d^2\frac{1}{Q^{12}}\frac{2^{\frac{43}{2}}}{\sqrt{\pi}r_0^3}
\left[\sum\limits_{j}C_{j}m_j^2\right]\left[\sum\limits_{j}D_{j}m_j^4\right]
\left(G_E^p(Q^2)+G_E^n(Q^2)\right)\;, \label{gnrq}
\end{equation}

\begin{equation}
G_M^{NR}\sim
\frac{1}{Q^8}\frac{2^{16}M_d}{\sqrt{\pi}r_0^3M}\left[\sum\limits_{j}C_{j}m_j^2\right]^2
\left(G_M^p(Q^2)+G_M^n(Q^2)\right)\;. \label{gnrm}
\end{equation}

The dimensional parameter $r_0$ can be found from the expression
for the deuteron matter radius in our deuteron model:
\begin{equation}
r_m^2=\frac{1}{4}\int_0^\infty(u_0^2(r)+u_2^2(r))r^2dr\;.
\label{rm}
\end{equation}
One can substitute wave functions of the form (\ref{funcukoord})
into this expression. So formula (\ref{rm}) specifies an
algebraic equation for $r_0$. Solution of this equation can be
found numerically.

It should be pointed out that main terms of expansion of charge
and magnetic form factors in (\ref{gnrc}) - (\ref{gnrm}) are
determined by $S$-state of deuteron only. The $D$-wave function
gives the contribution to the main term of the quadrupole from
factor. Its faster decrease at $Q^2\to\infty$ in comparison to
the other form factors is a consequence of a faster decrease of a
$D$-wave function at small distances in comparison to $S$-wave
(\ref{atorigin}). From the mathematical point of view the type of
leading terms in (\ref{gnrc})-(\ref{gnrm}) is a consequence of
conditions on the coefficients (\ref{cond}). The modification of
these conditions obviously results in change of the main terms in
(\ref{gnrc})-(\ref{gnrm}). From these formulas it is also noticed
that asymptotic expansions for the deuteron form factors contain
dependence on the asymptotics of nucleon form factors.

We emphasize, that in the other deuteron asymptotics
investigations only the power dependence on the transferred
momentum was calculated as a rule. In the present paper we give a
rigorous calculation of a multiplicative preasymptotical constant.

One can calculate relativistic asymptotics of form factors by
analogy with nonrelativistic case. For this calculation we use
the formulas (\ref{relnorm}), (\ref{anywfrel}), (\ref{cond}). As
a result we obtain:
\begin{equation}
G_{C,M}^R(Q^2)\;\sim\;\frac{Q^3}{2^{\frac{7}{2}}M^3}\,
G_{C,M}^{NR}(Q^2)\;, \label{gcm}
\end{equation}
\begin{equation}
G_{Q}^R(Q^2)\;\sim\;\frac{Q^4}{2^{\frac{11}{2}}M^4}\,
G_{Q}^{NR}(Q^2)\;. \label{gq}
\end{equation}
Notice that asymptotic expansions (\ref{gnrc})-(\ref{gnrm}) and
(\ref{gcm}),(\ref{gq}) are obtained for the first time in our
work. It is interesting to compare  obtained asymptotic
estimations (\ref{gnrc}) - (\ref{gnrm}), (\ref{gcm}),(\ref{gq})
with observable behavior of the deuteron characteristics. At
present time there exists the experimental information about
function $A(Q^2)$ entered the differential cross section of the
elastic $ed$-scattering. This function is expressed in terms of
the deuteron form factors \cite{GiG02}. The values of function
$A(Q^2)$ are known up to $Q^2\;\simeq\;$ 6 (GeV/c)$^2$.   For the
comparison with experimental data one needs to specify
asymptotics of the nucleon form factors.  It is naturally to
choose for nucleon form factors the asymptotic which is predicted
by the quark model \cite{GiG02} $G_M^{p,n}\sim {1}/{Q^4}$. Under
these conditions the power dependence on $Q^2$ of the function
$A(Q^2)$ coincides with experimentally observed one. The physical
consequences will be examined in detail in the other paper.

\section{Conclusion}

The theorem defining asymptotics of multiple integrals of some
special type is proved. With help of the proven theorem the
asymptotic expansion of the deuteron electromagnetic form factors
at $Q^2\;\to\;\infty$ is calculated for the first time. The
expansion is represented as a convergent series on in inverse
powers of momentum transfer. The asymptotic of the form factors
is found in terms of the conventional representation of the
deuteron wave function as a discrete su\-per\-po\-si\-ti\-on of
Yukawa-type terms. The asymptotic behavior of the form factors is
calculated in the nonrelativistic impulse approximation and in
the relativistic invariant impulse approximation proposed by the
authors in the instant form of the Poincare-invariant quantum
mechanics previously. It is established that relativistic
corrections change the power dependence of the form factors on the
momentum transfer at $Q^2\;\to\;\infty$ and slow down its
decrease. It is also found that relativistic effects result in
the agreement of the theoretical asymptotics and the
experimentally observed behavior of the structure function
$A(Q^2)$ at highest achieved momentum transfers.

\section{Acknowledgments}

This work was supported in part by Russian Foundation of Basic
Researches (grant 07-02-00962).

\end{document}